\documentclass{article}
\usepackage{spconf,amsmath,graphicx}

\usepackage{amssymb}
\usepackage{latexsym}
\usepackage{threeparttable}
\usepackage{booktabs}
\usepackage{algorithm}
\usepackage{algorithmic}
\usepackage{bm}
\usepackage{multirow}
\usepackage{float}
\usepackage{url}
\usepackage{xcolor}
\usepackage{array}
\usepackage{amsmath}
\usepackage{makecell}

\allowdisplaybreaks[4]

\title{The USTC-NERCSLIP Systems for the ICMC-ASR Challenge}
\name{\begin{tabular}{c} Minghui Wu$^{1,\dag}$\thanks{$^\dag$Contributed to this work equally. This work was supported by the National Natural Science Foundation of China (62101523) and the joint AI laboratory of CMB-USTC (FTIT2022058).}, Luzhen Xu$^{1,\dag}$, Jie Zhang$^{1,*}$\thanks{$^*$Corresponding author (jzhang6@ustc.edu.cn).}, Haitao Tang$^{2}$, Yanyan Yue$^{2}$, Ruizhi Liao$^{2}$,\\ Jintao Zhao$^{2}$,  Zhengzhe Zhang$^{1}$, Yichi Wang$^{1}$, Haoyin Yan$^1$, Hongliang Yu$^{2}$, Tongle Ma$^{2}$, \\ Jiachen Liu$^{2}$, Chongliang Wu$^{2}$, Yongchao Li$^{2}$, Yanyong Zhang$^{1}$, Xin Fang$^{2}$, Yue Zhang$^{3}$\end{tabular}}

\address{$^{1}$University of Science and Technology of China, China \\ 
$^{2}$Anhui iFLYTEK Universal Language Technology, China; 
$^{3}$Huawei Technology, China}
\begin{document}
%
\maketitle
\begin{abstract}
This report describes the submitted system to the In-Car Multi-Channel Automatic Speech Recognition (ICMC-ASR) challenge, which considers the ASR task with multi-speaker overlapping and Mandarin accent dynamics in the ICMC case. We implement the front-end speaker diarization using the self-supervised learning representation based multi-speaker embedding and beamforming using the speaker position, respectively. For ASR, we employ an iterative pseudo-label generation method based on fusion model to obtain text labels of unsupervised data. To mitigate the impact of accent, an Accent-ASR framework is proposed, which captures pronunciation-related accent features at a fine-grained level and linguistic information at a coarse-grained level. On the ICMC-ASR eval set, the proposed system achieves a CER of 13.16\% on track 1 and a cpCER of 21.48\% on track 2, which significantly outperforms the official baseline system and obtains the first rank on both tracks.

\end{abstract}
\begin{keywords}ICMC-ASR challenge, speaker diarization, multi-channel beamforming, pseudo-label.
\end{keywords}
\section{System description}
\label{sec:intro}

\textbf{Front-end processing:}
For the front-end processing, we employ the guided source separation (GSS) framework~\cite{wan2023ustc}, but with several modifications, including channel selection and beamforming. We propose a multi-source sound localization module using energy and phase differences to select the channel for the target speaker. Compared to the traditional maximum signal-noise ratio (SNR) selection criterion, it can alleviate the impact of non-target speakers with high speaking volume. The recursive smoothing technique is exploited for the estimation of  power spectral density matrices in the design of MVDR beamformer, which is shown to be more effective to suppress interference sources and noises and provide higher-quality single-channel audio  for the downstream ASR.



\begin{table}[htbp]
 \centering
 \renewcommand{\thetable}{1}
 \caption{Training Data Description.}
 \resizebox{0.90\linewidth}{!}{
 \setlength{\tabcolsep}{1.0 mm}
 \fontsize{8.5}{9.5}\selectfont
 \vspace{3mm}
 \begin{tabular}{c|c|c|c}
  \toprule
  \textbf{Dataset}  & \textbf{Data type} & \textbf{Label type} & \textbf{Hours}\\
  \hline
  ICMC & Far+Near+speed$\times$3 & \multicolumn{1}{c|}{\multirow{2}{*}{supervised}} & 386 \\
  \cline{1-2}
  \cline{4-4}
  ICMC+addNoise & Far+speed$\times$3 & \multicolumn{1}{c|}{} & 1544 \\
  \hline
  3D-SPEAKER & Near & \multicolumn{1}{c|}{\multirow{5}{*}{unsupervised}} & 1124 \\
  \cline{1-2}
  \cline{4-4}
  AliMeeting & Far+Near & \multicolumn{1}{c|}{} & 236 \\
  \cline{1-2}
  \cline{4-4}
  AISHELL-4 & Far+Near & \multicolumn{1}{c|}{} & 240 \\
  \cline{1-2}
  \cline{4-4}
  Aidatatang & Near & \multicolumn{1}{c|}{} & 200 \\
  \cline{1-2}
  \cline{4-4}
  MagicData & Near & (text is not & 180 \\
  \cline{1-2}
  \cline{4-4}
  KeSpeech & Near &  allowed to use) & 1542 \\
  \cline{1-2}
  \cline{4-4}
  \multicolumn{1}{c|}{\multirow{3}{*}{WenetSpeech}} & Near & \multicolumn{1}{c}{} & \multicolumn{1}{|c}{\multirow{3}{*}{5500}} \\
  \multicolumn{1}{c|}{} & (drama, talk, & \multicolumn{1}{c}{} & \multicolumn{1}{|c}{}\\
  \multicolumn{1}{c|}{} & interview) & \multicolumn{1}{c}{} & \multicolumn{1}{|c}{}\\
  \hline
  \bottomrule
  
 \end{tabular}}
 \vspace{-2.0em}
\end{table}

\textbf{Data resource:}
All data resources used for the proposed system are summarized in Table 1, which include two categories. One originates from the official ICMC-ASR labelled data~\cite{wang2024icmc}, and the other from the external openSLR unlabelled data following the official rules~\cite{kawakami2020learning}. For the far-field data, the single-channel audio is extracted by oracle diarization and GSS. For unsupervised data, we propose an iterative pseudo-label generation (PLG) method based on the fusion ASR model. To enhance the generalization of the fusion model, we consider different input features and encoder structures: 1) conformer encoder-decoder (ED) model based on self-supervised learning representation (SSLR), and 2) ebranchformer ED model using Fbank. The former is employed to adapt long-duration audio, while the latter is used for short-duration counterparts. The SSLR is extracted by the adaptive wavLM model~\cite{wan2023ustc}, which is trained using the WenetSpeech full 2.5wh dataset. For pseudo-label generation, a small amount of supervised data $B_{0}$ is initially used to train and fuse into $PLG_{0}$. The unsupervised data is then split into different batches, say $B_{1}, ..., B_{N}$. Subsequently, the pseudo-labels for the unsupervised data in $B_{N}$ are generated by $PLG_{N-1}$ in a cyclic iterative fashion, and $PLG_{N}$ is trained and fused by $B_{N}$. As such, we can obtain 10,952 hours of data for ASR training in total.

\textbf{Speaker Diarization:}
The speaker diarization system trained on 1544 hours data in Table 1 consists of three components: multi-channel voice activity detection (VAD), clustering-based speaker diarization (CSD) and multi-channel target-speaker VAD (MC-TS-VAD)~\cite{wan2023ustc}. Due to the additional non-target speaker (recorder) in the official data, traditional i-vector can not effectively distinguish between target and non-target speakers. To improve the generalization of speaker embeddings, we consider the fusion of the SSLR-based x-vector for the i-vector  in the MC-TS-VAD module. This  embedding fusion enables the diarization system to fully exploit speaker information and is thus helpful for speaker separation.


\textbf{Speech Recognition:}
\begin{figure}[tbp] 
\centering 
\includegraphics[width=0.49\textwidth]{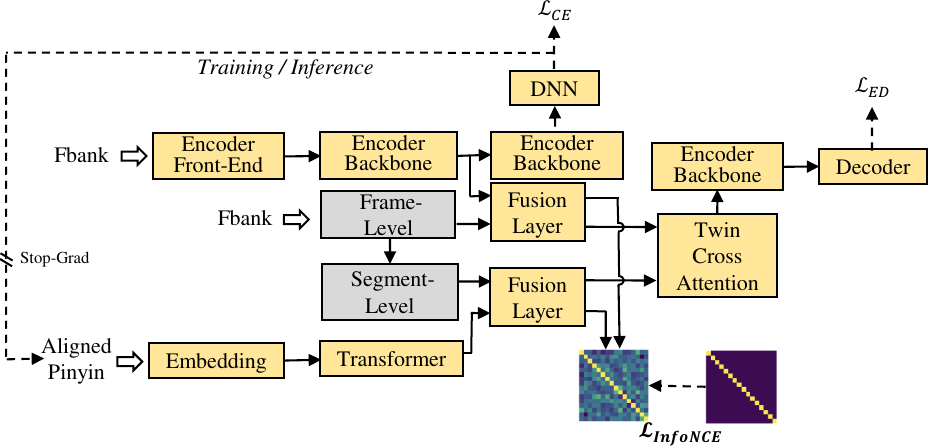} 
\vspace{-0.3cm}
\caption{The proposed Accent-ASR framework.}
\vspace{-0.3cm}
\label{accent}
\end{figure}
Due to the impact of accents on Mandarin ASR, we propose an Accent-ASR framework, which enables fine-grained units to capture pronunciation-related accent characteristics and coarse-grained units to learn linguistic information~\cite{shao2023decoupling}. As shown in Fig. 1, the proposed Accent-ASR is based on multi-task learning for aligning Pinyin sequences, and combines aligned Pinyin features with acoustic features of encoder through twin cross-attention fusion and contrastive learning, such that fine-grained units can better capture pronunciation details. Note that the aligned Pinyin features are derived from embedding and transformer modules. In the fusion layer, in addition to the aforementioned two types of features, frame-segment level speaker information is further introduced to help distinguish coarse-grained units that are generated by speakers of different accents. 
Besides, in order to increase the diversity of the Accent-ASR fusion model, we modify and combine the front and back sub-modules of encoder, where the front-end module includes Conv2D, VGG and gateCNN, and the backbone module consists of conformer and ebranchformer~\cite{zhang2022ustc}. But the decoder always adopts the transformer structure. 



\section{Performance Evaluation}
\label{sec:intro}
\textbf{Results on the Dev and Eval Sets of Track 1:}
The CER performance of the pseudo-label iterative generation based on the PLG fusion model is shown in Table 2. As the number of iterations increases, the performance of PLG significantly improves. Additionally, we observe that the wavLM-based ED model has competitive results given a relatively small amount of data. However, as the data amount  increases to 5000 hours, wavLM-based ED model becomes inferior to that of Fbank-based ED model.
Table 3 shows the results of ASR single and fusion models on Track 1 dev and eval sets. M0 represents the official baseline. M1-M5 are single Accent-ASR model composed of different encoder front-end and back-end sub-modules. M6 denotes the post fusion based on weight adaptation, which achieves a relative improvement of 52.8\% and 49.8\% compared to M0 on dev and eval sets, respectively.

\textbf{Results on the Dev and Eval Sets of Track 2:}
Table 4 presents the automatic speaker diarization and recognition (ASDR) results on Track 2 dev and eval sets. Our speaker diarization system achieves a diarization error rate (DER) of 10.21\%. In comparison to the official baseline, our system achieves a cpCER of 16.31\% on the dev set, which relatively improves by 75.3\%. Note that there is a difference of smaller than 5\% compared to Track 1 with oracle diarization. On the eval set, our system outperforms the official baseline with a relative improvement of 70.5\%.

\begin{table}[tbp]
 \centering
 \renewcommand{\thetable}{2}
 \caption{Pseudo-Label Iterative Generation Results (CER\%).}
\resizebox{0.98\linewidth}{!}{%
\setlength{\tabcolsep}{1.0 mm}{
\fontsize{9}{10}\selectfont
 \begin{tabular}{c|c|c|c|c}
  \toprule
  \hline
  \multicolumn{1}{c|}{\multirow{2}{*}{\textbf{Iteration}}}  & \multicolumn{1}{c|}{\multirow{2}{*}{\textbf{Data(hrs)}}} & \textbf{wavLM+} & \textbf{conv2d+} & \multicolumn{1}{c}{\multirow{2}{*}{\textbf{Fusion}}}\\

 \multicolumn{1}{c|}{} & \multicolumn{1}{c|}{} & \textbf{conformer ED} & \textbf{ebranchformer ED} & \multicolumn{1}{c}{} \\
  \hline
  B0 & 396 & 29.77 & 30.64 & 27.8 \\
\hline
  B1 & 3910 & 25.32 & 26.4 & 23.77 \\
\hline
  B2 & 5452 & 22.94 & 21.9 & 22.87 \\
\hline
  B3 & 10952 & 22.94 (5452h) & 21.14 & 20.83 \\
  \hline
  \bottomrule
 \end{tabular}}}
  \vspace{-1.5em}
\end{table}
\begin{table}[tbp]
 \centering
 \renewcommand{\thetable}{3}
 \caption{Results of single/fusion models on Track 1 (CER\%).}
\resizebox{0.98\linewidth}{!}{%
\setlength{\tabcolsep}{1.0 mm}{
\fontsize{9}{10}\selectfont
 \tabcolsep=0.5cm
 \begin{tabular}{c|c|c|c}
  \toprule
  \hline
  \textbf{ID} & \textbf{Model based on Accent-ASR} &\textbf{Dev} &\textbf{Eval}\\
\hline
  M0 & Official Baseline & 32.92 & 26.24  \\
\hline
  M1 & conv2d+conformer & 20.87 & -  \\
\hline
  M2 & conv2d+ebranchformer & 20.68 & - \\
\hline
  M3 & VGG+ebranchformer & 20.31 & - \\
\hline
  M4 & gateCNN+conformer & 19.83 & - \\
\hline
  M5 & gateCNN+ebranchformer & 18.53 & 14.72 \\
\hline
  M6 & fusion models based on M1-M5 & 15.54 & 13.16 \\
  \hline
  \bottomrule
 \end{tabular}}}
  \vspace{-0.7em}
\end{table}
\begin{table}[!tbp]
 \centering
 \renewcommand{\thetable}{4}
 \vspace{-1em}
 \caption{The ASDR results on Track 2 (\%).}
\resizebox{0.98\linewidth}{!}{%
\setlength{\tabcolsep}{1.0 mm}{
\fontsize{9}{10}\selectfont
 \tabcolsep=0.5cm
 \begin{tabular}{c|c|c|c}
  \toprule
  \hline
 \textbf{Model} & \textbf{Metric} & \textbf{Dev} & \textbf{Eval}\\
  \hline
 Speaker Diarization & DER (\%) & 10.21 & - \\
  \hline
  M0 Offical Baseline & \multicolumn{1}{c|}{\multirow{2}{*}{cpCER (\%)}} & 65.90 & 72.88 \\
   \cline{1-1}
  \cline{3-4}
  M6 Fusion models & \multicolumn{1}{c|}{} & 16.31 & 21.48\\ 

  \hline
  \bottomrule
 \end{tabular}}}
  \vspace{-1.5em}
\end{table}



\bibliographystyle{IEEEbib}
\vspace{-0.75em}
\bibliography{strings,refs}

\begin{thebibliography}{1}

\bibitem{wan2023ustc}
Ruoyu Wang, Maokui He, et~al.,
\newblock ``The ustc-nercslip systems for chime-7 challenge,''
\newblock in {\em Proc. CHiME 2023}, 2023, pp. 13--18.

\bibitem{wang2024icmc}
He~Wang, Pengcheng Guo, et~al.,
\newblock ``Icmc-asr: The icassp 2024 in-car multi-channel automatic speech recognition challenge,''
\newblock {\em arXiv preprint arXiv:2401.03473}, 2024.

\bibitem{kawakami2020learning}
Kazuya Kawakami, Luyu Wang, et~al.,
\newblock ``Learning robust and multilingual speech representations,''
\newblock {\em arXiv preprint arXiv:2001.11128}, 2020.

\bibitem{shao2023decoupling}
Qijie Shao, Pengcheng Guo, et~al.,
\newblock ``Decoupling and interacting multi-task learning network for joint speech and accent recognition,''
\newblock {\em IEEE/ACM TASLP}, vol. 32, 2023.

\bibitem{zhang2022ustc}
Weitai Zhang, Zhongyi Ye, et~al.,
\newblock ``The ustc-nelslip offline speech translation systems for iwslt 2022,''
\newblock in {\em Proc. IWSLT}, 2022, pp. 198--207.

\end{thebibliography}
\end{document}